\documentclass[11pt]{article}
\usepackage{moriond,epsfig}
\def\Journal#1#2#3#4{{#1} {\bf #2}, #3 (#4)}
\def\edth{\;\raise1.0pt\hbox{$'$}\hskip-6pt\partial\;}
\def\baredth{\;\overline{\raise1.0pt\hbox{$'$}\hskip-6pt
\partial}\;}
\def\gsim{~\rlap{$>$}{\lower 1.0ex\hbox{$\sim$}}}
\def\lsim{~\rlap{$<$}{\lower 1.0ex\hbox{$\sim$}}}

\def\d{{\rm d}}

\def\PRD{{\em Phys. Rev.} D}
\def\ApJ{\bf ApJ}

\def\be{\begin{equation}}
\def\ee{\end{equation}}
\def\bea{\begin{eqnarray}}
\def\eea{\end{eqnarray}}

\begin{document}
\vspace*{4cm}
\title{CMB Quadrupole induced polarisation from clusters and
  filaments}
\author{ Guo-Chin Liu$\ ^{1,2}$, Antonio da Silva$\ ^1$, Nabila Aghanim$\ ^1$}
\address{1.Institut d'Astrophysique Spatiale - UMR-8617
Universit$\acute{e}$ Paris-Sud, b$\hat{a}$timent 121 F-91405 Orsay, France\\
2.Institute of Astronomy and Astrophysics, Academia Sinica. 7F of
Condensed Matter Sciences and Physics Department Building,
National Taiwan University. No.1, Roosevelt Rd, Sec. 4 Taipei 106,
Taiwan, R.O.C. }

\maketitle\abstracts{We present the estimates of the Cosmic Microwave
  Background (CMB) polarisation power spectrum from galaxy clusters
  and filaments using hydrodynamical simulations of large scale
  structure formation. We focus in the present study on the CMB
  Quadrupole induced polarisation, which is the dominate secondary
  polarisation effect, and give the E and B mode power spectra between
  $l \sim 560$ to $20000$.}

%----------------------------------------------------------------
% Introduction
%----------------------------------------------------------------
\section{Introduction}
After recombination, the CMB photons scatter off free electrons in
ionised matter,
such as the hot gas trapped in the potential wells of galaxy clusters.
The presence of the CMB temperature quadrupole induces a linear
polarisation in the scattered radiation.

Sunyaev and Zel'dovich~\cite{SZ} were the first to estimate the
level of polarisation in galaxy clusters. These authors pointed
out that in addition to the primary CMB quadrupole there are two
other sources of temperature quadrupole seen by a cluster: a
quadrupole due to the transverse peculiar velocity of the cluster
 and double scattering. Studying the induced CMB polarisation due
to clusters can open a whole new window for cosmology. Sunyaev and
Zel'dovich proposed to use the polarisation to estimate cluster's
transverse velocity. Measuring the polarisation towards distant
clusters should provide us with an opportunity to observe the
evolution of the CMB quadrupole~\cite{SS}. Moreover, CMB
quadrupole seen by the clusters contains statistical information
on the last scattering surface of at the cluster position.
Therefore, measuring the cluster polarisation should help us to
beat the cosmic variance~\cite{KL,JP}.

In this work, we study the polarisation induced by galaxy clusters.
We focus in the statistical properties of such a polarised signal by
computing its angular power spectrum at small angular scales.  We
explore the signal associated with the primary CMB temperature
quadrupole since it dominates at the scales we are interested
in~\cite{SS}. To do so, we follow the method of Liu et al.~\cite{Liu},
in which the authors computed the polarised signal at reionisation
using N-body simulations in combination with an analytical description
to model the gas distribution.  Based on the same formulae and similar
approach, we estimate the polarisation from the hot gas in galaxy
clusters and the warm gas in filaments using hydrodynamical
simulations of large scale structure (LSS), which directly
account for the gas dynamics.

In Section 2 we describe the method of this work. We briefly review
the formulae of the polarisation power spectra and the hydrodynamical
simulations used in this paper. In Section 3, we present the results
of our work and give a discussion and conclusion in Section 4.

\section{Background}

The polarisation signal is usually described by the Stokes parameters
which can be combined to obtain a divergence free component, the
so-called $E$ mode, and a curl component, the so-called $B$ mode.

The power spectra of polarisation of E and B modes are obtained by
integrating the product of the visibility function $g$ and the
polarisation source.  In the case of primary polarisation, the source
is the quadrupole of the primary CMB temperature anisotropies. This is
a first order signal. When dealing with second order effects, one has
to take into account possible modulations of the visibility function
by the electron density perturbations. This was done in Liu et
al.~\cite{Liu}. More specifically, the power spectra were found to
be:

\begin{equation}
C_{(E,B)l}=(4\pi)^2\frac{9}{16}\frac{(l+2)!}{(l-2)!} \sum_m\int
k^2 \d k \left\langle \left |\int \d\tau g(\tau) \delta_e({\bf
k},\tau) Q^{(m)}({\bf k},\tau) T_{(E,B)l}^{(m)}(kr)\right |^2
\right\rangle,
\label{DEL2}
\end{equation}
where $\delta_e({\bf k},\tau)$ is the electron density perturbation as
a function of scale $\bf k$ and conformal time $\tau$, and
$Q^{(m)}=\sqrt{\frac{4\pi}{5}}\int\d^3 {\bf k} \Delta^{(m=0)}_{T2}
({\bf k})Y_2^{m*}(\hat{\bf k})$ is the projected scalar mode of
primary CMB quadrupole ($\Delta^{(m=0)}_{T2}$) in the frame of $\delta_e$.
The visibility function $g(\tau)$ has a
simple physical meaning, being the probability that a photon had its
last scattering at epoch $\tau$ and reaches the observer at the
present time, $\tau_0$. It is defined as
$g(\tau)\equiv -\frac{\d\kappa}{\d\tau}{\rm
e}^{\kappa(\tau_0)-\kappa(\tau)}$,
where $\kappa ({\tau})$ is the electron-scattering optical depth at
time $\tau$. The term $T_{(E,B)l}^{(m)}(kr)$ is given by the
combination of the spherical Bessel functions (for details see Table
1 in Liu et al.~\cite{Liu2}).

In order to calculate the polarisation anisotropy spectra using the
formulae mentioned above, we compute the visibility function and the
electron density distribution from clusters and filaments using
non-radiative hydrodynamical simulations of a $\Lambda$CDM cosmology,
with matter density parameter, $\Omega_{{\rm m}}=0.3$, cosmological
constant density, $\Omega_{\Lambda}=0.7$, hubble parameter, $h=0.71$,
and baryon density, $\Omega_b=0.044$. The initial density field
was constructed using $160^3$ particles of both baryonic and dark
matter, perturbed from a regular grid of fixed comoving size $L=100
h^{-1} {\rm Mpc}$. The CDM matter power spectrum used in the
simulation had a shape parameter, $\Gamma=0.21$, and normalization
$\sigma_8=0.9$. The run was started at redshift $z=50$ and evolved to
$z=0$ using the publicly available {\tt Hydra} (AP$^3$M/SPH)
code~\cite{CTP,PC}. The gravitational softening was fixed at
$25\,h^{-1} {\rm kpc}$ in physical units below $z=1$, and above this
redshift held constant to $50\,h^{-1} {\rm kpc}$ in comoving co-ordinates.

In the simulation, the gas is assumed ionised if the temperature is
higher than $10^5$ K. We further define two gas phases based on the
baryon collapsed density: when the gas overdensity, $\delta$, is
greater than the density contrast at collapse, $\Delta_c = 178
\Omega_m(z)^{-0.55}$, the gas is in clusters; the low density phase
defined by $5<\delta<\Delta_c$ encompasses a warm gas which we call
filaments. The visibility function is obtained from the time evolution
of the ionisation fraction in our run for $T>10^5$ K.  For the
computation of $C_{(E,B)l}$, we evaluate the electron density
fluctuations on a regular grid with $600^3$ cells inside the comoving
box, which corresponds to a fixed cell separation of $0.17\, h^{-1}
{\rm Mpc}$ in comoving coordinates.

Finally, the only remained unknown term is the quadrupole of the
temperature anisotropies. Given the cosmological parameters and the
ionisation fraction from our simulation, the publicly available code
CMBFast~\cite {ZS} can output the time evolution of the temperature
quadrupole component with different ${\bf k}$-basis.  Note that the
perturbations of tensor mode are such that the normalizations to COBE
and to $\sigma_8=0.9$, used in our simulation, are consistent.
We assume that the tensor spectral index satisfies the single field
inflationary consistency condition $n_t=-r/8$ with $r$ being the
ratio of tensor to scalar quadrupole $r=C^{T}_{l}/C^{S}_{l}$. Here
$r=0.05$ and $n_t=-0.00625$ are applied when running CMBFast.

\section{Power spectra of polarisation of clusters and filaments}

\begin{figure}
%\rule{5cm}{0.2mm}\hfill\rule{5cm}{0.2mm}
%\vskip 2.5cm
%\rule{5cm}{0.2mm}\hfill\rule{5cm}{0.2mm}
\hskip 2.5cm
\psfig{figure=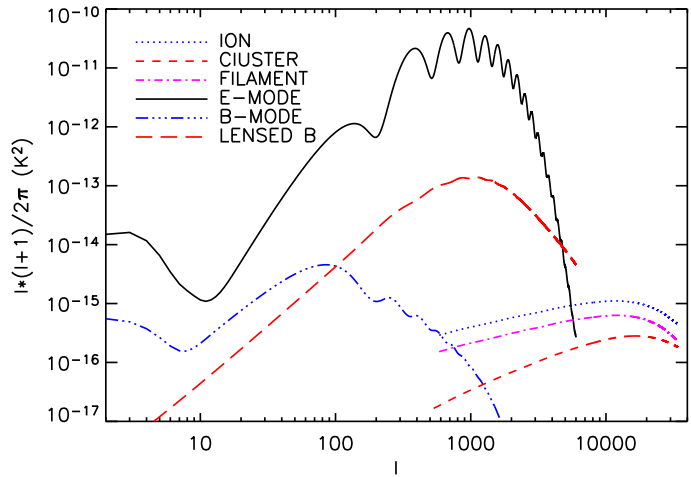,height=7cm}
\caption{Polarisation power spectra for all ionized particles,
filaments and cluster are shown by dotted, dot-dashed and short dashed
curves, respectively. We also plot the primary $E$ mode induced by
scalar perturbation (solid line), primary $B$ mode (triple dot-dashed
line) from tensor mode and $B$ mode from lensed $E$ mode (long dashed
line).}
\label{cl_all}
\end{figure}

We now investigate the contribution to the polarisation power spectra
from the two different gas phases defined previously.  We show the
results in Fig({\ref{cl_all}}). Note that the range of multipoles $l$
is limited by the resolution (for high $l$s) and box size (for low
$l$s) of the simulation.
The highest multipole $l$ is set by the corresponding angular size of
$0.17\,h^{-1} {\rm Mpc}$ at $z=0.72$ in our simulation.

We get the same power for the secondary E and B mode polarisation with
a maximum difference is less than $10^{-6}$. The reason for the
equality of the total E- and B-mode power spectra is essentially that
the first-order quadrupole whose scattering produces the polarisation
is dominated by large scales, and so has a random orientation relative
to the small-scale fluctuations in the electron density. Scattering of
the quadrupole by the small-scale fluctuations therefore on average
excites E- and B- modes equally~\cite{Hu}~\cite{Liu}.

We now compare the polarisation signal from two phases: the cluster
and so-called filaments. The power spectrum from all the ionized
particles (with $T>10^5$ K) is also shown in the figure
({\ref{cl_all}}). The two curves for all ionized particles and
filaments have similar shapes peaking at $l \sim 10^4$.  The power
spectrum of polarisation in clusters also peaks at the same scales but
shows much less power on the large scales. The power spectrum of the
filaments is about one order of magnitude larger than that of the
clusters at $l \sim 1000$. This is due to the fact that the dominant
contributions comes from the filaments on large scales. At smaller
angular scales, the power spectrum of filaments is still larger than
that of clusters. Recall the equation(\ref{DEL2}), the power spectra
is the integral of electron density weighted by the visibility
function. From our simulation, the visibility function peaks at $z
\gsim 3$ and its full width half maximum is bounded by $z \sim 1.2$
and $z \sim 6$. In this range, the power of electron density for
filaments is quite larger than for clusters. In the other words, most
of the scattering producing the linear polarisation occurs for $z
\gsim 1.2 $ when the relative contribution of the clusters is still
small (for details see Liu et al.~\cite{Liu2}).

\section{Discussion and Conclusion}

Using hydrodynamical LSS simulations, we compute the
$E$ and $B$ mode polarisation power spectra of galaxy clusters and
filaments due to the primary CMB quadrupole. As expected, these power
spectra dominate at small angular scales. The question is then: will
the secondary polarisation signal, associated with clusters and
filaments, be a problem for the measurement of the primary $E$ and $B$
mode power spectra? To compare the polarisation spectra from the
different physical processes at different scales, we show, in addition to
our computations, the primary $E$ and $B$ modes by scalar and tensor
mode perturbations respectively. We also show in the same figure the
expected $B$ mode signal from the power conversion of $E$ mode by
gravitational lensing. In our results, the $E$ mode signal from
clusters and filaments dominates the primary signal at $l > 6000$.
Below this scale, the secondary signal is always very small
and therefore is not expected to significantly contaminate the primary
plarisation signal.

For the $B$ mode polarisation, and although some uncertainties in
amplitude of the tensor mode perturbations, our secondary $B$ mode
signal is definitively not a problem for finding the gravitational
wave induced polarisation at scales $l \lsim 100$. However, the
gravitational lensing-induced signal which peaks around $l \sim
1000$ is a major issue~\cite{ZS2}. The power spectrum of the $B$
mode polarisation from clusters and filaments is two orders of
magnitude smaller than the lensing and it dominates at smaller
angular scales. It might be observed at these scales if the
lensing-induced $B$ modes can be significantly cleaned, see for
example Seljak U. and Hirata (2004) \cite{SH}.

\end{document}